\documentstyle[12pt]{article}
\begin{document}
\newcommand{\siml}{\stackrel{<}{\sim}}
\newcommand{\simg}{\stackrel{>}{\sim}}
\newcommand{\lleq}{\stackrel{<}{=}}
\baselineskip=1.333\baselineskip

\noindent
\begin{center}
{\large\bf

Nonextensive thermodynamics of the
two-site Hubbard model
} 
\end{center}

\begin{center}
Hideo Hasegawa
\footnote{e-mail:  hasegawa@u-gakugei.ac.jp}
\end{center}

\begin{center}
{\it Department of Physics, Tokyo Gakugei University  \\
Koganei, Tokyo 184-8501, Japan}
\end{center}
\begin{center}
{\rm (January 1, 2005)}
\end{center}
\thispagestyle{myheadings}

\begin{center} 
{\bf Abstract}   \par
\end{center} 
Thermodynamical properties
of canonical and grand-canonical ensembles
of the half-filled two-site Hubbard model
have been discussed
within the framework of the nonextensive statistics (NES).
For relating the physical temperature $T$ to the Lagrange
multiplier $\beta$, two methods have been adopted:
$T=1/k_B \beta$ in the method A 
[Tsallis {\it et al.} Physica A {\bf 261} (1998) 534], 
and  $T=c_q/k_B \beta$ in the method B
[Abe {\it et al.} Phys. Lett. A {\bf 281} (2001) 126],
where $k_B$ denotes the Boltzman
constant, $c_q= \sum_i p_i^q$, $p_i$ the 
probability distribution of the $i$th state,
and $q$ the entropic index.
Temperature dependences of specific heat and
magnetic susceptibility have been calculated 
for $1 \lleq q \lleq 2$,
the conventional Boltzman-Gibbs statistics being recovered 
in the limit of $q = 1$.
The Curie constant $\Gamma_q$ of the susceptibility
in the atomic and low-temperature limits 
($t/U \rightarrow 0, \; T/U \rightarrow 0$)
is shown to be given by $\Gamma_q=2\:q\: 2^{2(q-1)}$
in the method A, and $\Gamma_q=2\:q$ in the method B,
where $t$ stands for electron hoppings and $U$ intra-atomic 
interaction in the Hubbard model.
These expressions for $\Gamma_q$ are shown to agree with the results of 
a free spin model which has been studied also 
by the NES with the methods A and B.
A comparison has been made between the results for canonical and 
grand-canonical ensembles of the model.

\vspace{0.5cm}


\vspace{1.5cm}
\noindent
{\it Corresponding Author}

\noindent  
Hideo Hasegawa \\ 
Department of Physics, Tokyo Gakugei University \\ 
4-1-1 Nukui-kita machi, Koganei, Tokyo 184-8501, Japan \\
Phone: 042-329-7482, Fax: 042-329-7491 \\
e-mail: hasegawa@u-gakugei.ac.jp



\newpage

\section{INTRODUCTION}


In the last several years, 
there is an increased interest in the
nonextensive statistics (NES), which was initially
proposed by Tsallis \cite{Tsallis88,Tsallis98,Tsallis04}.
This is because the standard method based on
the Boltzman-Gibbs statistics (BGS) cannot properly deal with
{\it nonextensive} systems where the physical quantity 
associated with $N$ particles 
is not proportional to $N$ \cite{Note3}.
The nonextensivity has been realized in three different classes of
systems: (a) systems with long-range interactions,
(b) small-scale systems with fluctuations of temperatures
or energy dissipations, and 
(c) multifractal systems \cite{Tsallis04,NES}.
In a gravitating system with the long-range interaction,
which is a typical case (a), the specific heat 
becomes negative \cite{Padman90}. 
A cluster of 147 sodium atoms, which belongs to the case (b), 
has been reported to show
the negative specific heat \cite{Schmidt01}:
note that the specific heat is never negative in the 
canonical BGS \cite{Note1}.

Tsallis \cite{Tsallis88}\cite{Tsallis98}
has proposed a generalized entropy in the NES
defined by

\begin{equation}
S_q= k_B \left( \frac{\sum_i p_i^q-1}{1-q} \right),
\end{equation}
where $k_B$ denotes the Boltzman constant,
$p_i$ the probability distribution of the system
in the $i$th configuration, and
$q$ the entropic index. 
The entropy of BGS, $S_{BG}$, is obtained from Eq. (1)
in the limit of $q = 1$, as given by
\begin{equation}
S_1 = S_{BG}= - k_{B} \sum_i p_i \;{\rm ln} \:p_i.
\end{equation}
The nonextensivity in the Tsallis entropy is satisfied as follows.
Suppose that the total system is divided into two 
independent subsystems with the probability distributions,
$p_i^{(1)}$ and $p_i^{(2)}$.
The total system is described by the factorized
probability distribution $p_{ij}=p_i^{(1)}\:p_j^{(2)}$.
The entropy for the total system given by Eq. (1) satisfy the relation:
\begin{equation}
S_q=  S_q^{(1)}+S_q^{(2)}
+\left(\frac{1-q}{k_{B}}\right) S_q^{(1)} S_q^{(2)},
\end{equation}
where $S_q^{(k)}$ stands for the entropy of the $k$th
subsystem. Equation (3) shows that the entropy is extensive
for $q=1$ and nonextensive for $q \neq 1$:
the quantity $\mid q-1 \mid$ expresses the measure of
the nonextensivity.
The new formalism has been successfully
applied to a wide range of nonextensive systems
including physics, chemistry, mathematics, astronomy, 
geophysics, biology, medicine, economics, 
engineering, linguistics, and others \cite{NES}.

The current NES, however, is not complete,
having following unsolved issues.

\noindent
(i) A full and general understanding of the relation
between the entropic index $q$ and the underlying
microscopic dynamics is lacking.
The index $q$ is usually obtained 
in a phenomenological way by a fitting of experimental
or computational available data.
It has been reported that
the observed velocity distribution
of galaxy clusters significantly deviates from 
BGS distribution, which may be fitted well by the NES distribution
with $q \sim 0.23$ \cite{Lavagno98}.
The index $q$ in self-gravitating systems has been shown to be
$q = 0.60 \sim 0.92$ for $n= 3 \sim 20$
with the use of the relation: $n=1/(1-q)+1/2$ for 
the entropic index $q$ and the polytrope index $n$
given by $P(r) \propto \rho(r)^{1+1/n}$, where
$P(r)$ is pressure and $\rho(r)$ mass density \cite{Taruya03}.
However, the index $q$ has been successfully determined 
{\it a priori} in some cases \cite{Tsallis04}.
In one-dimensional dissipative map,
the index is given by $q=0.2445$ \cite{Lyra98} from its scaling
property of dynamic attractors.
In nanometric systems consisting of $N$ noninteracting
particles, the NES distribution is shown to
arise from fluctuating $\beta$ whose distribution is given by
the $\chi^2$ (or $\Gamma$) distribution, 
leading to $q = 1+2/N$ \cite{Wilk00,Beck02,Raja04}.

\noindent
(ii) The second issue is that it is not clear
how to relate the physical temperature $T$
to the introduced Lagrange multiplier $\beta$.
So far two methods have been proposed:
\begin{eqnarray}
T &=& \frac{1}{k_B \beta},
\hspace{2cm}\mbox{(method A)}\\
&=& \frac{c_q}{k_B \beta},
\hspace{2cm}\mbox{(method B)}
\end{eqnarray}
where $c_q= \sum_i p_i^q$.
The method A proposed in Ref. \cite{Tsallis98} is the same 
as the extensive BGS.
The method B is introduced so as to satisfy the {\it zero}th law
of thermodynamical principles and the generalized Legendre
transformations \cite{Abe01}.
It has been demonstrated that the negative specific heat 
of a classical gas model which is realized in the method A \cite{Abe99},
is remedied in the method B \cite{Abe01}.
Results calculated with the use of 
the two methods have been compared in self-gravitating systems 
\cite{Sakagami03}.

In this paper, we wish to apply the NES
to the Hubbard model, which is one of the
most important models 
in condensed-matter physics (for a recent review
on the Hubbard model, see Ref.\cite{Note2}).
The Hubbard model consists of the tight-binding term
expressing electron hoppings and the short-range interaction
between two electrons with opposite spins.
The model provides us with good
qualitative description for many interesting phenomena
such as magnetism, electron correlation, and superconductivity.
In particular, the model has been widely employed 
for a study of metallic
magnetism. 
In the atomic limit where the electron interaction is
much larger than electron hoppings,
the Hubbard model with the half-filled band reduces to
a local-spin model such as the Heisenberg and Ising models.
Despite the simplicity of the Hubbard, however, 
it is very difficult to
obtain its exact solution. 
In order to get a reasonable solution, various technical
methods have been developed \cite{Note2}.
It is possible to employ a small number of lattice models
to obtain an analytical solution.
The two-site Hubbard model is employed as a simple model
which can be analytically solved.
Thermodynamical and magnetic properties of the two-site model
have been studied within 
the BGS \cite{Suezaki72,Shiba72,Bernstein74}.
Although the two-site Hubbard model is a toy model,
it is an exactly solvable quantum system.
Actually this model may be adopted to describe 
interesting phenomena in real
systems such as organic charge-transfer salts
with dimerized structures \cite{Suezaki72}-\cite{Kozlov96}.
Because the two-site Hubbard model is considered to
belong to the case (b) of small-scale systems mentioned above, 
it is worthwhile to apply the NES to the model
in order to investigate nonextensive effects 
on its thermodynamical and magnetic properties, 
which is the purpose of the present paper.

The paper is organized as follows. After discussing
the NES for grand-canonical ensembles (GCE)
of the two-site Hubbard model in Sec. 2, we will calculate
the specific heat, susceptibility and the Curie
constant of the susceptibility
in the atomic limit calculated by the NES
with the two methods A and B for the $T-\beta$ relations
given by Eqs. (4) and (5). 
In Sec. 3, we present calculations
for canonical ensembles (CE) of the model, which are
compared to those for GCE.
The final Sec. 4
is devoted to discussions and conclusions.
In appendix A, free spin systems have been discussed
also by using the NES with the two methods A and B.

\section{Grand-canonical ensembles}

\subsection{Entropy, energy and free energy}

We consider GCE
of the half-filled, two-site Hubbard model given by
\begin{equation}
\hat{H}= -t \sum_{\sigma}
( a_{1\sigma}^{\dagger} a_{2\sigma} 
+  a_{2\sigma}^{\dagger} a_{1\sigma}) 
+ U \sum_{j=1}^2 n_{j \uparrow} n_{j \downarrow }
- h \sum_{j=1}^2 (n_{j \uparrow} - n_{j \downarrow}),  
\end{equation}
where $n_{j\sigma}
= a_{j\sigma}^{\dagger} a_{j\sigma}$,
$a_{j\sigma}$ denotes the annihilation operator of an electron with
spin $\sigma$ on a site $j$ (=1, 2), 
$t$ the hopping integral,
$U$ the intraatomic interaction and $h$ an applied magnetic
field in an appropriate unit.
Eigen values of the system are given by \cite{Bernstein74}
\begin{eqnarray}
\epsilon_i&=& 0,
\hspace{4cm}\mbox{for $i=1$ ($n_i=0$)} \\
&=& \pm t \pm h,  
\hspace{3cm}\mbox{for $i=2$ to 5 ($n_i=1$)} \\
&=& 0, \pm 2h, U, U/2 \pm \Delta, 
\hspace{1cm}\mbox{for $i=6$ to 11 ($n_i=2$)} \\
&=& U\pm t \pm h,  
\hspace{3cm}\mbox{for $i=12$ to $15$ ($n_i=3$)} \\
&=& 2U,  
\hspace{4cm}\mbox{for $i=16$ ($n_i=4$)}
\end{eqnarray}
where $\Delta=\sqrt{U^2/4+4 t^2}$, and $n_i$
expresses the number of electrons in the $i$th state.
The grand-partition function in BGS, $\Xi_{BG}$,
is given by ($k_B=1$ hereafter) \cite{Bernstein74}
\begin{eqnarray}
\Xi_{BG}&=& Tr \:e^{-(\hat{H}-\mu \hat{N})/T}, \\
&=&1+4 \;{\rm cosh}(h/T) \:{\rm cosh} (t/T) \:
e^{\mu/T}\nonumber \\
&&+ [1+2 \:{\rm cosh}(2h/T) + {\rm e}^{-U/T}
+ 2 \:{\rm e}^{-U/2T} \:{\rm cosh} (\Delta/T)] 
e^{2\mu/T} \nonumber \\
&&+ 4 \;{\rm cosh}(h/T) \:{\rm cosh}(t/T) \:e^{-(U-3\mu)/T}
+ e^{-2(U-2\mu)/T}.
\end{eqnarray}
Here {\it Tr} stands for trace, 
$\hat{N}=\sum_{i=1}^2 \sum_{\sigma} n_{i\sigma}$,
and the chemical potential is $\mu=U/2$ 
independent of the temperature in the half-filled case
where the number of total electrons is $N_e=2$.
By using the standard method in the BGS, 
we can obtain various thermodynamical quantities of the system 
\cite{Bernstein74}.

Now we adopt the NES in which the entropy $S_q$
is given by \cite{Tsallis88}\cite{Tsallis98}
\begin{equation}
S_q=\left( \frac{Tr \:(\hat{\rho}_q^q) - 1}{1-q} \right).
\end{equation}
Here 
$\hat{\rho}_q$ denotes the generalized density matrix,
whose explicit form will be determined shortly [Eq. (18)].
We impose the three constraints given by 
\begin{eqnarray}
Tr \:(\hat{\rho}_q)&=&1, \\
\frac{Tr \:(\hat{\rho}_q^q \: \hat{H})}{Tr \:(\hat{\rho}_q^q)}
&\equiv& <\hat{H}>_q = E_q, \\
\frac{Tr \:(\hat{\rho}_q^q \: \hat{N})}{Tr \:(\hat{\rho}_q^q)}
&\equiv& <\hat{N}>_q = N_q,
\end{eqnarray}
where the normalized formalism is adopted \cite{Tsallis98}.  
The variational condition for the entropy with
the three constraints given by Eqs. (15)-(17)
yields
\begin{equation}
\hat{\rho}_q=\frac{1}{X_q} {\rm exp}_q 
\left[-\left( \frac{\beta}{c_q} \right) 
(\hat{H}-\mu \hat{N}-E_q+\mu N_q) \right],
\end{equation}
with
\begin{eqnarray}
X_q&=&Tr\: \left( {\rm exp}_q 
\left[ -\left( \frac{\beta}{c_q} \right) 
(\hat{H}-\mu \hat{N}-E_q+\mu N_q) \right] \right), \\
c_q&=& Tr \:(\hat{\rho}_q^q) = X_q^{1-q},
\end{eqnarray}
where 
${\rm exp}_q (x) \equiv [1+(1-q)x]^{\frac{1}{1-q}}$
is the generalized exponential function.
Lagrange multipliers $\beta$ and $\mu$ relevant to
the constraints given by Eqs. (16) and (17) are given by 
the relations: 
\begin{eqnarray}
\beta&=&\frac{\partial S_q}{\partial E_q},\\
\mu &=& -\frac{1}{\beta}\frac{\partial S_q}{\partial N_q}.
\end{eqnarray}
The entropy $S_q$ in Eq. (14) is expressed by
\begin{equation}
S_q= \left( \frac{X_q^{1-q}-1}{1-q} \right)
\equiv  \ln_q \; (X_q),
\end{equation}
where ${\rm ln}_q (x) \equiv (x^{1-q}-1)/(1-q)$
is the generalized logarithmic function.

In relating the physical temperature $T$ 
to the Lagrange multiplier $\beta$,
we have adopted the two methods \cite{Tsallis98}\cite{Abe01}:
\begin{eqnarray}
T &=& \frac{1}{\beta},
\hspace{2cm}\mbox{(method A)}\\
&=& \frac{c_q}{\beta}.
\hspace{2cm}\mbox{(method B)}
\end{eqnarray}

In the limit of $q=1$, we get the results obtained 
in the BGS:
$\hat{\rho}_{BG}={\rm e}^{-\beta (\hat{H}-\mu \hat{N})}/\Xi_{BG}$, 
$E_{BG} = Tr \: (\hat{\rho}_{BG} \hat{H})$,
$N_{BG} = Tr \: (\hat{\rho}_{BG} \hat{N})$,
$X_{BG}={\rm e}^{\beta (E_{BG}-\mu N_{BG})}\;\Xi_{BG}$,
$\Xi_{BG}=Tr \: ({\rm e}^{-\beta (\hat{H}-\mu \hat{N})})$,
and $S_{BG} = - Tr\: (\hat{\rho}_{BG} \;{\rm ln} \:\hat{\rho}_{BG})$.

It is necessary to point out that 
$E_q$ and $X_q$ have to be determined self-consistently
by Eqs. (16)-(20) with $\mu$ determined by Eq. (22)
for $N_q=N_e$ and a given temperature $T$ 
because they are mutually dependent.
In the half-filled case, however, calculations become easier
because $\mu=U/2$ independent of the temperature.
The calculation of thermodynamical quantities
in the NES generally
becomes more difficult than that in BGS.
In our numerical calculations to be reported in this paper,
simultaneous equations for $E_q$ and $X_q$
given by Eqs. (16)-(20) are solved
by using the Newton-Raphson method.
The iteration have started with initial values of $E_q$ and $X_q$
obtained from the BGS ($q=1$).
Numerical calculations have been made
for $1 \lleq q \lleq 2$ which is appropriate
for nanoscale systems\cite{Wilk00,Beck02,Raja04}.

Figures 1(a)-1(f) show the temperature dependence of 
the energy $E_q$ of GCE calculated for $h=0$.
Bold solid curves in Fig. 1(a), 1(b) and 1(c) show
$E_1$ in the BGS
calculated for $U/t=0$, 5 and 10, respectively.
The ground-state energy at $T=0$ is $E_1/t=$ -2.0, -0.70156 and -0.38516
for $U/t=0$, 5 and 10, respectively.
With increasing $q$ value above unity, the gradient
of $E_q$ is much decreased in the method A, as shown in Figs. 1(a)-1(c).
This trend is, however, much reduced in the method B, as Figs. 1(d)-1(f)
show. 
This behavior is more clearly seen in the temperature dependence
of the specific heat $C_q$, as will be discussed shortly
[Figs. 3(a)-3(f)].

Temperature dependences of the entropy 
for $h=0$ are plotted in Figs. 2(a)-2(f).
Figures 2(a), 2(b) and 2(c) express $S_q$
for $U/t=0$, 5 and 10, respectively,
calculated by the method A, and Figs. 2(d)-2(f)
those calculated by the method B.
Bold curves denote the results for the BGS, where the entropy 
is quickly increased at low temperature when the interaction is increased.
When the $q$ value is more increased above unity, $S_q$ 
is more rapidly increased
at very low temperatures and its saturation value 
at higher temperatures becomes smaller.
This behavior is commonly realized in the results calculated
by the methods A and B.
A difference between the two results is ostensibly small
because $S_q$ shows a saturation at low temperatures.

\subsection{Specific heat}

First we consider the specific heat, which is given by
\begin{equation}
C_q= \left( \frac{d \beta}{d T} \right)
\left( \frac {d E_q}{d \beta} \right).
\end{equation}
Because $E_q$ and $X_q$ are determined by
Eqs. (16)-(20), we get simultaneous equations for
$d E_q/d \beta$ and  $d X_q/d \beta$, given by
\begin{eqnarray}
\frac {d E_q}{d \beta} 
&=& a_{11}  \left( \frac {d E_q}{d \beta} \right)
+ a_{12} \left( \frac {d X_q}{d \beta} \right) + b_1, \\
\frac {d X_q}{d \beta} 
&=& a_{21}  \left( \frac {d E_q}{d \beta} \right)
+ a_{22} \left( \frac {d X_q}{d \beta} \right),
\end{eqnarray}
with
\begin{eqnarray}
a_{11}&=& q \beta X_q^{q-2} 
\sum_i w_i^{2q-1} \epsilon_i,  \\
a_{12}&=& -X_q^{-1} E_q
-\beta q (q-1) X_q^{q-3} \sum_i w_i^{2q-1}
\epsilon_i (\epsilon_i-\mu n_i -E_q+ \mu N_e), \\
a_{21}&=& \beta X_q^q,\\
a_{22}&=& 0,\\
b_1&=& - q X_q^{q-2} \sum_i w_i^{2q-1}
\epsilon_i (\epsilon_i-E_q),\\
w_i &=& <i\mid {\rm exp}_q 
\left [- \left( \frac{\beta}{c_q}\right) 
(H-\mu N - E_q + \mu N_q) \right] \mid i>, \nonumber \\
&=& \left[1-(1-q) \left( \frac{\beta}{c_q} \right)
(\epsilon_i -\mu n_i- E_q + \mu N_q) \right]^{\frac{1}{1-q}},\\
X_q &=& \sum_i w_i.
\end{eqnarray}
The specific heat is then given by
\begin{equation}
C_q= \left( \frac{d \beta}{d T} \right) 
\left( \frac{b_1}{1-a_{11}-a_{12}a_{21}} \right).
\end{equation}
with
\begin{eqnarray}
\frac{\partial \beta}{\partial T}
&=& - \beta^2, 
\hspace{7cm} \mbox{(method A)} \\
&=& - \left( \frac{\beta^2}
{X_q^{1-q} - \beta (1-q) X_q^{-q} 
(d X_q/d \beta) }\right), 
\hspace{1cm} \mbox{(method B)} \\
\frac{d E_q}{d \beta} &=& \frac{b_1}
{(1-a_{11}-a_{12}a_{21})},\\
\frac{d X_q}{d \beta} &=& \frac{a_{21}b_1}
{(1-a_{11}-a_{12}a_{21})},
\end{eqnarray}

It is worthwhile to examine the limit of $q = 1$
of Eqs.(29)-(35), which reduce to
\begin{eqnarray}
a_{11}&=& \beta E_1,  \\
a_{12}&=& -X_1^{-1} E_1, \\
a_{21}&=& \beta X_1,\\
b_1&=& - <\epsilon_i^2>_1 + <\epsilon_i>_1^2.
\end{eqnarray}
yielding
\begin{equation}
C_1= \beta^2 (<\epsilon_i^2>_1 - <\epsilon_i>_1^2),
\end{equation}
with
\begin{eqnarray}
<Q_i>_1&=&X_1^{-1} \sum_i 
e^{-\beta (\epsilon_i-\mu n_i-E_1+\mu N_e)} \:Q_i, \\
E_1 &=& <\epsilon_i>_1, \\
X_1&=&\sum_i 
e^{-\beta (\epsilon_i-\mu n_i-E_1+\mu N_e)}.
\end{eqnarray}
The expression for $C_1$ agrees with $C_{BG}$
obtained in the BGS.

Figures 3(a)-3(f) show the specific heat of GCE calculated
for $h=0$. 
$C_1$ in BGS for $U/t=0$ shown by the bold solid curve 
in Fig. 3(a), has a peak at $T/t \sim 0.65$.
Figure 3(c) shows that
for $U/t=10$, this peak splits into two.
A lower peak arises from low-lying collective
spin-wave-like excitations while higher one
from single-particle excitations \cite{Shiba72}\cite{Bernstein74}.
For intermediate $U/t=5$ these two peaks overlap [Fig. 3(b)]. 
The temperature
dependences of the specific heat $C_q$
calculated with the use of the method A
for $U/t=0$, 5 and 10 
are plotted in Figs. 3(a), 3(b) and 3(c), respectively.
We note that when $q$ is larger than unity, peaks become broader.
Figures 3(d), 3(e) and 3(f) show the temperatures
dependence of the specific heat $C_q$ 
calculated by the method B
for $U/t=0$, 5 and 10, respectively.
Although general property of the $q$ dependence of 
the specific heat of the method B is similar 
to that of the method A, the effect of the nonextensivity
in the method B becomes smaller than that in the method A.

\subsection{Susceptibility}

In the NES,
the magnetization induced by an applied field $h$ is obtained by
\begin{eqnarray}
m_q&=& < \mu_i>_q,
\end{eqnarray}
leading to the susceptibility given by
\begin{eqnarray}
\chi_q&=& \frac{\partial m_q}{\partial h}\mid_{h=0},\\
&=& -E_q^{(2)}+\beta^{-1} X_q^{-q} X_q^{(2)},
\end{eqnarray}
where $\mu_i = - \:\partial \epsilon_i/\partial h$ and
$E_q^{(2)}=\partial^2 E_q/\partial h^2\mid_{h=0}$ {\it et al}.
With calculations using Eqs. (16)-(20),
we get simultaneous equations for 
$E_q^{(2)}$ and $X_q^{(2)}$, given by
\begin{eqnarray}
E_q^{(2)}&=&a_{11} E_q^{(2)}+ a_{12} X_q^{(2)} + f_1, \\
X_q^{(2)}&=&a_{21} E_q^{(2)}+ a_{22} X_q^{(2)} + f_2, 
\end{eqnarray}
with
\begin{eqnarray}
f_1&=& -2 \:\beta \:q \:X_q^{q-2} \sum_i w_i^{2q-1}\: \mu_i^2,\\
f_2&=& \beta^2 \:q \:X_q^{2(q-1)} \sum_i w_i^{2q-1} \:\mu_i^2,
\end{eqnarray}
where $a_{ij}$ ($i,j=1,2$) are given by Eqs. (29)-(31).
The susceptibilities $\chi_q$ in the methods A and B 
are expressed by
\begin{eqnarray}
\chi_q&=& \left(\frac{-a_{12}+\beta^{-1}X_q^{-q}(1-a_{11})}
{1-a_{11}-a_{12}a_{21}} \right)\;f_2
=\frac{f_2}{a_{21}},
\end{eqnarray}
with the $T-\beta$ relations given by Eqs. (24) and (25), 
respectively.

In the limit of $q = 1$, Eqs. (54) and (55) reduce to
\begin{eqnarray}
f_1&=& -2 \:\beta \:\left<\mu_i^2\right>_1,\\
f_2&=& \beta^2 \:X_1 \:\left< \mu_i^2\right>_1,
\end{eqnarray}
leading to the susceptibility in the BGS:
\begin{equation}
\chi_{BG}=\chi_1=\beta <\mu_i^2>_1,
\end{equation}
where the bracke $<\cdot>_1$ is given by Eq. (46).

The BGS susceptibility for $U/t=0$ has a peak at
$T/t \sim 0.65$ as Fig. 4(a) shows.
With increasing $U/t$, the magnitude of $\chi_{BG}$
is enhanced by the interaction,
and its peak positon becomes lower \cite{Shiba72}\cite{Bernstein74},
as Figs. 4(b) and 4(c) show:
the horizontal scale of Fig. 4(c) is
enlarged compared with Figs. 4(a) and 4(b).
The temperature dependences of the susceptibility $\chi_q$
calculated by the method A for $U/t=0$, 5 and 10 
are plotted in Figs. 4(a), 4(b) and 4(c), respectively.
We note that as increasing $q$ above unity, the peak
in $\chi_q$ becomes broader.
Figures 4(d), 4(e) and 4(f) show the temperature
dependence of the susceptibility $\chi_q$
calculated by the method B 
for $U/t=0$, 5 and 10, respectively.
Again the effect of the nonextensivity in the
method B becomes smaller than that in the method A.

\subsection{Curie constant in the atomic limit}

The half-filled Hubbard model with $t/U \ll 1$
reduces to a local-spin model with
the superexchange interaction $J \sim t^2/U$.
In the limit of $t/U \rightarrow 0$ (atomic limit) 
for which $J \rightarrow 0$,
the susceptibility of the two-site Hubbard model in BGS is given by
\begin{eqnarray}
\chi_{BG}&=& \frac{2}{T (1+ e^{-U/2T})}, \\
&=& \frac{2}{T}, 
\hspace{2cm}\mbox{for $T \ll U$} \\
&=& \frac{1}{T}.
\hspace{2cm}\mbox{for $T \gg U$}
\end{eqnarray}
Defining the effective, temperature-dependent 
Curie constant $\Gamma_q(T)$ by
\begin{equation}
\Gamma_q(T)=T\:\chi_q(T),
\end{equation}
we note from Eqs. (61) and (62)
that it varies from $\Gamma_{1}(0)=2$
for localized moments to $\Gamma_{1}(\infty)=1$
for delocalized moments.

By using the NES, we get
the Curie constant of our model in the low-temperature limit 
($T \ll U$), given by
\begin{eqnarray}
\Gamma_q (0) &=& 2\:q\:M^{q-1}=2\: q \: 2^{2(q-1)}, 
\hspace{1cm}\mbox{(method A)} \\
&=& 2\:q,
\hspace{4cm}\mbox{(method B)}
\end{eqnarray}
with
\begin{equation}
X_q=M =2^2,
\end{equation}
where $M$ denotes the number of states with
the lowest value of $\epsilon_i - \mu n_i$.
Similarly the Curie constant in the high-temperature
limit ($T \gg U$) is given by
\begin{eqnarray}
\Gamma_q (\infty) &=& q \:M_{\infty}^{q-1}=q \: 4^{2(q-1)}, 
\hspace{1cm}\mbox{(method A)} \\
&=& q,
\hspace{4cm}\mbox{(method B)}
\end{eqnarray}
with
\begin{equation}
X_q=M_{\infty}=4^2,
\end{equation}
where $M_{\infty}$ expresses the number
of available states in our model.
Expressions given by Eqs. (64) and (65) 
are consistent with the results
for free spin systems calculated by NES 
with the methods A and B [Eqs. (90) and (91)],
whose detail is discussed in appendix A.

Solid curves in Fig. 5(a) and 5(b) show 
the temperature dependence of
the inversed susceptibility $1/\chi_q$ 
for $t=0$ with various $q$ values
calculated by the methods A and B, respectively.
We notice that as increasing the $q$ value,
the gradient of $1/\chi_q$ is decreased in both the methods.  
Solid curves in Fig. 5(c) and 5(d) express the temperature
dependence of $\Gamma_q$ calculated by the methods A and B,
respectively.
We note that $\Gamma_1$ is 2 at $T/U \sim 0$ 
and approaches 1 at $T/U \sim 10$, as Eqs. (61) and (62) show.
$\Gamma_2(T)$ in
the method A is 16 at $T/U \sim 0$ and it is rapidly increased
as increasing the temperature with
the maximum at $T/U \sim 4$, above which it is decreased.
In contrast, $\Gamma_2(T)$ in the method B starts
from 4 at $T/U\sim 0$, and approaches 2 at $T/U=10$ with a peak at
$T/U \sim 0.2$.

We note that in Figs. 5(a)-5(d) that the Curie constant
is increased with increasing $q$. This is more clearly
seen in Figs. 5(e) and 5(f), where we plot
the $q$ dependences of $\Gamma_q(T)$
calculated by the methods A and B, respectively.
Circles in Fg. 5(e) 
show $\Gamma_q(T)$ at $T=0.02$
calculated by the method A, which nicely obeys 
the relation $\Gamma_q(0)=2\:q \:2^{2(q-1)}$ [Eq. (64)].

On the contrary, $\Gamma_q(\infty)$ in the model A is
expressed by squares in Fig. 5(e) where the chain curve
denotes $\Gamma_q(\infty) = q \:4^{2(q-1)}$ [Eq. (67)].
A disagreement between the chain curve and the result
of $T/U=10$ arises from a fact
that the temperature of $T/U=10$ does not
correspond the high-temperature limit,
which is realized at $T/U > 100$.
Circles in Figure 5(f) show the $q$ dependence of
$\Gamma_q(T)$ at $T/U=0.02$ calculated by the method A,
which follows the relation $\Gamma_q(0)=2 q$ given by Eq. (65).
On the contrary, $\Gamma_q(T)$ at $T/U=10$
calculated by the method B is plotted
by squares in Fig. 5(f), approximately
following the chain curve 
given by $\Gamma_q(\infty)=q$ [Eq. (68)].

\section{Canonical ensembles}

In the previous section, we have discussed thermodynamical
properties of GCE of the two-site Hubbard model.
It is straightforward to extend our study
to CE of the model.
The canonical partition function $Z_{BG}$ in the BGS is
given by \cite{Suezaki72}
\begin{equation}
Z_{BG}=1+2 \:{\rm cosh}(2h/T) + e^{-U/T}
+ 2 \:e^{-U/2T} \:{\rm cosh} (\Delta/T).
\end{equation}

The generalized density matrix $\hat{\rho}_q$
for CE in the NES may be determined 
by the variational condition for the
entropy given by Eq. (14) with the {\it two} constraints
given by Eqs. (15) and (16) \cite{Hasegawa04}:
\begin{equation}
\hat{\rho}_q=\frac{1}{X_q} {\rm exp}_q 
\left[-\left( \frac{\beta}{c_q} \right) 
(\hat{H}-E_q) \right],
\end{equation}
with
\begin{eqnarray}
X_q&=&Tr\: \left( {\rm exp}_q 
\left[ -\left( \frac{\beta}{c_q} \right) 
(\hat{H}-E_q) \right] \right), \\
c_q&=& Tr \:(\hat{\rho}_q^q) = X_q^{1-q},
\end{eqnarray}
where $\beta$ stands for the Lagrange multiplier and
$E_q$ is given by Eq. (16).
The $T-\beta$ relations in
the methods A and B are given by Eqs. (24) and (25), respectively.
The specific heat is expressed by Eqs. (29)-(40) but with
$a_{12}$, $w_i$ and $X_q$ given by \cite{Hasegawa04}
\begin{eqnarray}
a_{12}&=& -X_q^{-1} E_q
-\beta q (q-1) X_q^{q-3} \sum_i w_i^{2q-1}
\epsilon_i (\epsilon_i-E_q), \\
w_i &=& \left[1-(1-q) \left( \frac{\beta}{c_q} \right)
(\epsilon_i- E_q ) \right]^{\frac{1}{1-q}},\\
X_q &=& \sum_i w_i,
\end{eqnarray}
where a sum over $i$ (= 6 to 11) is only for $n_i=2$ in Eq. (9).
The susceptibility is similarly
expressed by Eqs. (54)-(56) with $a_{12}$, $w_i$ and $X_q$ 
given above. 
The Curie constants with $t/U=0$
in the low-temperature limit are 
given by  Eqs. (64) and (65) in the methods A and B, 
respcitively, while those in the high-temperature limit 
are given by Eqs. (67) and (68)
but with $q \rightarrow (4/3)\:q$ and $M_{\infty}=C^4_2$
where $C^n_k=n!/k!(n-k)!$.

Temperature dependences of the specific heat $C_q$ of CE
calculated for $h=0$ are plotted in Figs. 6(a)-6(f).
Bold curves in Figs. 6(a)-6(c) show the temperature 
dependence of $C_{BG}$ in the BGS.
For $U/t=0$, $C_{BG}$ calculated for CE
is similar to that for GCE
shown in Fig. 3(a).
For $U/t=10$, however, the result for CE
is rather different from that for GCE
shown in Fig. 3(c):
the latter has a broad peak at $T/t > 1$
while the former not.
The high-temperature peak in $C_{BG}$ of GCE
is responsible to charge fluctuations
whose effect is included in the GCE treatment
but not in the CE one. The temperature dependences of
the specific heat in the NES calculated by the method A 
are plotted in Figs. 6(a)-6(c), whereas those calculated 
by the method B are shown in Figs. 5(d)-5(f).

Figures 7(a)-7(c) show the temperature dependence of
the susceptibility in the NES calculated by the method A.
Results calculated by the method B are plotted in
Figs. 7(d)-7(f).
The difference between the results of CE and
GCE is small.
This is because the susceptibility is not sensitive to
charge fluctuations.
Effects of the nonextensivity on $C_q$ and $\chi_q$
in the method B 
become smaller than those in the method A for CE, 
just as for GCE.

Although results calculated for CE and GCE coincide
for infinite systems, it is not the case for finite ones.
The CE-method is appropriate
for an analysis of real nanoscale materials
where the number of total electrons is conserved.
In contrast, the GCE-method is expected to be better 
in guessing properties of infinite systems 
from results of finite systems \cite{Shiba72}. 
This is understood as follows.
Suppose that an infinite system is divided into many
segments. Charge fluctuations exist between adjacent
segments, which are taken into account
in the GCE statistics. 

\section{Discussions and Conclusions}

Recent progress in atomic engineering makes it possible
to create nanoscale materials 
with the use of various methods
(for reviews, see Ref. \cite{Note4}).
Nanomagnetism shows interesting properties different from 
bulk magnetism.
Nanoclusters consisting of transition metals such as
${\rm Fe}_N$ ($N$=15-650) \cite{Heer90}, 
${\rm Co}_N$ ($N$=20-200) \cite{Bucher91},
and ${\rm Ni}_N$ ($N$=5-740) \cite{Apsel96} have been synthesized 
by laser vaporization and
their magnetic properties have been measured,
where $N$ denotes the number of atoms per cluster.
Magnitudes of magnetic moments are increased 
with reducing $N$ \cite{Apsel96}.
It is shown that magnetic moments in Co monatomic chains
constructed on Pt substrates
are larger than those in monolayer Co
and bulk Co \cite{Gambardella02}. 
Recently Au nanoparticles with average diameter of 1.9 {\it nm}
(including 212 atoms), which are protected
by polyallyl amine hydrochloride (PAAHC),
are reported to show ferromagnetism while 
bulk Au is diamagnetic \cite{Yamamoto04}.
This is similar to the case of gas-evaporated Pd fine particles
with the average diameter of 11.5 {\it nm}
which show the ferromagnetism whereas bulk Pd is paramagnetic 
\cite{Shinohara03}. 
The magnetic property of
4 Ni atoms with the tetrahedral structure
in magnetic molecules of metallo-organic substance
$[{\rm Mo}_{12}{\rm O}_{30}(\mu_2-{\rm OH})_{10}
{\rm H}_2\{{\rm Ni}({\rm H}_2{\rm O}_3) \}_4] 
\cdot 14 {\rm H}_2{\rm O}$ has been studied \cite{Postnikov04}.
Their temperature-dependent susceptibility
and magnetization process
have been analyzed by using the Heisenberg model
with the antiferromagnetic 
exchange couplings between Ni atoms \cite{Postnikov04}.
Similar analysis has been made for six-, eight-, ten- and
twelve-membered Fe atoms
in a family of magnetic molecules \cite{Gatteshi00}.

It is well known that
the Hubbard model is more suitable
for a study on bulk magnetism
of transition metals than the Heisenberg or 
Ising model which is best applied to rare-earth-metal 
magnets like Gd \cite{Note2}. 
This is true also for transition-metal nanomagnetism.
It is interesting to make a comparison of our result 
obtained for the two-site Hubbard model 
with experimental data of nanoclusters
consisting of two transition-metal atoms. 
Our NES calculations have shown that
the results for $q=2$, which is realized for a two-atom 
molecule if we assume $q=1+2/N$
\cite{Wilk00,Beck02,Raja04}, are rather different from
those for $q=1$ (BGS). 
Unfortunately {\it ideal} clusters 
including only two transition-metal atoms 
are not available at the moment. 
Some charge-transfer salts
like tetracyanoquinodimethan (TCNQ)
with dimerized structures, have been analyzed
by using the two-site Hubbard model within the BGS \cite{Bernstein74}.
Their susceptibility and specific heat 
were studied by taking into account
the effect of the interdimer hopping, which is smaller than 
the intradimer hopping, 
by a perturbation to the two-site Hubbard model \cite{Bernstein74}.
For an analysis of nanochains including multiple
dimmers within the NES,
it is necessary to know the size dependence of
the $q$ value appropriate 
for them \cite{Hasegawa05}.


For a better understanding of nanomagnetism,
we are requested to perform NES calculations 
of the $N$-site Hubbard model with a larger $N$
in one, two and three dimensions.
The number of eigen values is $4^N$ for the GCE and
$C^{2N}_{N}$ for the CE with the half-filled 
electron occupancy.
When the orbital degeneracy is taken into account
for a more realistic description, these numbers become
$4^{ND}$ and $C^{2ND}_{ND}$ for GCE and CE,
respectively, where $D$ ($=2L+1$) denotes the orbital degeneracy
for the orbital quantum number $L$ ($L=2$ for 3d transition metals)
\cite{Note5}.
Realistic NES calculations for larger nanosystems
become much laborious.
We hope that by extending the current engineering
technique, it would be possible to synthesize ideal,
small nanoclusters
consisting of transition-metal atoms.

To summarize, 
we have discussed thermodynamical and magnetic properties 
of GCE and CE of the two-site Hubbard model
within the frame work of the NES.
This is the first application of NES to thermodynamical
and magnetic properties of the Hubbard model, as far as
the author is aware of. 
The temperature dependences of the specific heat 
and susceptibility calculated by the two methods A and B
for $T-\beta$ relation, are qualitatively the same, 
and they change significantly when $q$ deviates from unity. 
The two methods, however, yield a quite different 
$q$ dependence of the Curie constant, which is 
demonstrated for the atomic-limit case, and which is
consistent with the result
for free spin models as discussed in appendix A.
Although the result of the method B is considered to be
more reasonable than that of the method A,
we reserve to conclude which of the two methods
is correct for the two-site Hubbard model 
at this stage.
It is necessary to further develop the NES theory
in order to clarify the two issues (i) and (ii) raised 
in the introduction.

\section*{Acknowledgements}
This work is partly supported by
a Grant-in-Aid for Scientific Research from the Japanese 
Ministry of Education, Culture, Sports, Science and Technology.  


\vspace{1cm}
\noindent
{\large\bf Appendix A$\;\;$  NES for a free spin model }

We consider Hamitonian expressing
the $N$ Ising spins with $S=1/2$, given by
\begin{equation}
H= - g\:h \sum_{j=1}^N S_{zj} \equiv -g\:h S_z, 
\end{equation}
where $g \:(=2)$ denotes the $g$ factor and
$h$ an applied magnetic field in an
appropriate units.
Eingen states are classified by
$m_s=S_z$, which runs from $-N/2$ to $N/2$.
The multiplicity of the state specified by $m_s$
is given by \cite{Reis02}
\begin{equation}
f(m_s,N)=\frac{N!}{(N/2+m_s)! \:(N/2-m_s)!},
\end{equation}
which satisfies the sum rule:
\begin{equation}
\sum_{m_s=-N/2}^{N/2} f(m_s,N)=2^N\equiv M.
\end{equation}
Hereafter we specify the state with the index $i$, which
runs from 1 to $M$. In the case of $N=2$, for example,
we get $m_i=-1$ for $i=1$, $m_i=0$ for $i=2,3$
and $m_i=1$ for $i=4$.

By employing the NES for CE as in Sec. 3, we get the
probability density given by 
\begin{eqnarray}
p_i &=& \frac{w_i}{X_q}, \\
X_q &=& \sum_i w_i, \\
w_i &=& {\rm exp}_q \left[\left(\frac{\beta}{c_q}\right)
(h m_i+E_q) \right], \\
c_q &=& \sum_i p_i^q = X_q^{1-q}, \\
E_q &=& X_q^{-1} \sum_i h m_i\;
\left( {\rm exp}_q \left[\left(\frac{\beta}{c_q}\right)
(h m_i+E_q) \right] \right)^q, 
\end{eqnarray}
The thermal average of the magnetization is given by
\begin{equation}
m_q = g\: \left(\frac{\sum_i p_i^q \: m_i}
{\sum_i p_i^q} \right), 
\end{equation}
from which we get the susceptibility given by
\begin{eqnarray}
\chi_q &=& \frac{\partial m_q}{\partial h} \mid_{h=0}, \\
&=& g^2 \left( \frac{q \beta}{X_q^{1-q}} \right)
\: \frac{1}{X_q} \sum_i w_i^{2q-1} m_i^2,\\
&=& \frac{N q \beta}{c_q}. 
\end{eqnarray}
In deriving Eq. (88), we have adopted
$w_i=1$ and $X_q=2^N$, and
\begin{equation}
2^{-N} \sum_{m_s=-N/2}^{N/2} f(m_s,N) \;m_s^2 = \frac{N}{4},
\end{equation}
for $(\beta/c_q)(h \:m_i+E_q) \ll 1$ with $h \rightarrow 0$.
When we adopt the method A given by $T=1/\beta$ [Eq. (24)], 
Eq. (88) becomes
\begin{eqnarray}
\chi_q &=& \frac{\Gamma_q}{T},
\end{eqnarray}
with the Curie constant $\Gamma_q$ given by
\begin{eqnarray}
\Gamma_q &=& N \:q\:  M^{q-1}=N\: q \: 2^{N(q-1)},
\end{eqnarray}
which agrees with the result of Ref. \cite{Mar00} 
previously obtained 
by using the Tsallis' normalized scheme \cite{Tsallis98}.
Equation (91) has an anomalous exponential dependence on $N$,
and shows {\it dark magnetism}: the apparent number of spins
is larger than the actual one \cite{Mar00,Portesi95}.
On the contrary, when we adopt the method B given by 
$T = c_q/\beta$ [Eq. (25)], the Curie constant becomes
\begin{eqnarray}
\Gamma_q &=& N \:q,
\end{eqnarray}
whose $N$ dependence seems reasonable.

By using the $T-\beta$ relation given by
\begin{eqnarray}
\frac{1}{T} = \frac{\beta}{c_q+ (1-q) \beta E_q},
\end{eqnarray}
Reis {\it et al.} \cite{Reis02} have obtained 
the result same as Eq. (91).
This is due to a fact that
with $E_q = 0$ for $h=0$,
Eq. (92) reduces to $T=c_q/\beta$ which is nothing but 
the $T-\beta$ relation in the method B.

When we adopt the relation: $q=1+2/N$ obtained for nanoscale systems
\cite{Wilk00,Beck02,Raja04}, 
Eqs. (90) and (91) become
\begin{eqnarray}
\Gamma_q &=& 4 N \left( 1+\frac{2}{N} \right), 
\hspace{2cm}\mbox{(method A)} \\
&=& N \left( 1+\frac{2}{N} \right).
\hspace{2cm}\mbox{(method B)} 
\end{eqnarray}
Equation (94) is consistent with the result of the Ising
model with $N \rightarrow \infty$.


\newpage

\begin{figure}
\caption{
The temperature dependences of the energy $E_q$
of grand-canonical ensembles for $h=0$ 
with (a) $U/t=0$, (b) 5 and (c) 10 calculated by
the method A (GCE-A),
and those with (d) $U/t=0$, (e) 5 and (f) 10
calculated by the method B (GCE-B):
$q=1.0$ (bold solid curves), 1.1 (dotted curves), 
1.2 (dashed curves), 1.5 (chain curves) 
and 2.0 (solid curves).
}
\label{fig1}
\end{figure}

\begin{figure}
\caption{
The temperature dependences of the entropy $S_q$
of grand-canonical ensembles for $h=0$
with (a) $U/t=0$, (b) 5 and (c) 10 calculated by
the method A (GCE-A) for $h=0$,
and those with (d) $U/t=0$, (e) 5 and (f) 10
calculated by the method B (GCE-B):
$q=1.0$ (bold solid curves), 1.1 (dotted curves), 
1.2 (dashed curves), 1.5 (chain curves) 
and 2.0 (solid curves).
}
\label{fig2}
\end{figure}

\begin{figure}
\caption{
The temperature dependences of the specific heat $C_q$
of grand-canonical ensembles for $h=0$
with (a) $U/t=0$, (b) 5 and (c) 10 calculated by
the method A (GCE-A),
and those with (d) $U/t=0$, (e) 5 and (f) 10
calculated by the method B (GCE-B):
$q=1.0$ (bold solid curves), 1.1 (dotted curves), 
1.2 (dashed curves), 1.5 (chain curves) 
and 2.0 (solid curves).
}
\label{fig3}
\end{figure}

\begin{figure}
\caption{
The temperature dependences of the susceptibility $\chi_q$
of grand-canonical ensembles
for (a) $U/t=0$, (b) 5 and (c) 10 calculated by
the method A (GCE-A),
and those for (d) $U/t=0$, (e) 5 and (f) 10
calculated by the method B (GCE-B):
$q=1.0$ (bold solid curves), 1.1 (dotted curves), 
1.2 (dashed curves), 1.5 (chain curves) 
and 2.0 (solid curves).
}
\label{fig4}
\end{figure}

\begin{figure}
\caption{
The temperature dependence of $1/\chi_q$ 
in the atomic limit ($t/U=0$)
of grand-canonical ensembles 
for various $q$ values
calculated by (a) the method A (GCE-A) and (b) the method B (GCE-B).
The temperature dependence of $\Gamma_q(T)$ 
in the atomic limit ($t/U=0$)
for various $q$ values
calculated by (c) the method A and (d) B.
The $q$ dependence of the Curie constant $\Gamma_q(T)$ 
at $T/U=0$ and 10 calculated
by (e) the method A and (f) B.
Chain curves in (e) and (f) express 
$\Gamma_q(\infty)=q\: 4^{2(q-1)}$ and
$\Gamma_q(\infty)=q$, respectively (see text).
}
\label{fig5}
\end{figure}

\begin{figure}
\caption{
The temperature dependences of the specific heat $C_q$
of canonical ensembles for $h=0$ 
with (a) $U/t=0$, (b) 5 and (c) 10 calculated by
the method A (CE-A),
and those with (d) $U/t=0$, (e) 5 and (f) 10
calculated by the method B (CE-B):
$q=1.0$ (bold solid curves), 1.1 (dotted curves), 
1.2 (dashed curves), 1.5 (chain curves) 
and 2.0 (solid curves).
}
\label{fig6}
\end{figure}

\begin{figure}
\caption{
The temperature dependences of the susceptibility $\chi_q$
of canonical ensembles 
for (a) $U/t=0$, (b) 5 and (c) 10 calculated by
the method A (CE-A),
and those for (d) $U/t=0$, (e) 5 and (f) 10
calculated by the method B (CE-B):
$q=1.0$ (bold solid curves), 1.1 (dotted curves), 
1.2 (dashed curves), 1.5 (chain curves) 
and 2.0 (solid curves).
}
\label{fig7}
\end{figure}


\begin{thebibliography}{99}
\bibitem{Taruya03}
A. Taruya and M. Sakagami, Physica A {\bf 322} (2003) 285.
Phys. Rev. Lett. {\bf 90} (2003) 181101: In the latter paper,
it is shown that when Tsallis' old formalism is employed [Ref. 1],
the $q-n$ relation is given by $n=1/(q-1)+(3/2)$ leading to
$q=1.05 \sim 1.66$ for $n= 3 \sim 20$, where $n$ is the polytrope index.
\end{thebibliography}
\end{document}